\documentclass[%
 aip,
cp,  
 amsmath,amssymb,
 reprint,%
]{revtex4-2}

\usepackage{graphicx}
\usepackage{dcolumn}
\usepackage{bm}

\usepackage[utf8]{inputenc}
\usepackage[T1]{fontenc}
\usepackage{mathptmx}

\newcommand{\bb}{\mbox{$\beta\beta$}}
\newcommand{\nue}{\mbox{$\nu_e$}}
\newcommand{\ise}{\mbox{$^{82}$Se}}
\newcommand{\sitwo}{\mbox{$^{32}$Si}}
\newcommand{\ptwo}{\mbox{$^{32}$P}}
\newcommand{\pbten}{\mbox{$^{210}$Pb}}
\newcommand{\biten}{\mbox{$^{210}$Bi}}
\newcommand{\poten}{\mbox{$^{210}$Po}}
\newcommand{\ueight}{\mbox{$^{238}$U}}
\newcommand{\thtwo}{\mbox{$^{232}$Th}}
\newcommand{\dru}{\mbox{events per keV$_{\rm ee}$$\cdot$kg$\cdot$day}}
\newcommand{\um}{\textmu m}

\newcommand{\eve}{\mbox{eV$_{\rm ee}$}}

\begin{document}

\title{Background Rejection in Highly Pixelated Solid-State Detectors}

\author{Alvaro E. Chavarria} 
 \email[Corresponding author: ]{chavarri@uw.edu}
\affiliation{
  Center for Experimental Nuclear Physics and Astrophysics, University of Washington, Seattle, Washington 98195, United States
}

\date{\today} 

\begin{abstract}
Highly pixelated solid-state detectors offer outstanding capabilities in the identification and suppression of backgrounds from natural radioactivity. We present the background-identification strategies developed for the DAMIC experiment, which employs silicon 
charge-coupled devices to search for dark matter. DAMIC has demonstrated the capability to disentangle and measure the activities of every $\beta$ emitter from the \sitwo , \ueight\ and \thtwo\ decay chains in the silicon target. Similar strategies will be adopted by the Selena Neutrino Experiment, which will employ hybrid amorphous \ise/CMOS imagers to perform spectroscopy of \bb\ decay and solar neutrinos. We present the proposed experimental strategy for Selena to achieve zero-background in a 100-ton-year exposure.
\end{abstract}

\maketitle

\section{\label{sec:intro}Low-background highly pixelated solid-state detectors}

Highly pixelated detectors capture high-resolution images of the ionizing tracks of particles in a target, which provide a wealth of information on the type and origin of the particles.
The images are constructed from the ionization charge generated in the target, which is drifted by an electric field toward a pixel array instrumented for charge readout.
For solid-state targets, the devices can be either monolithic or hybrid.
Monolithic devices have the pixel array on the same silicon substrate as the ionization target, while hybrid devices couple a silicon pixelated readout chip to a separate ionization target.
For low-energy applications, charge sensors with extremely low pixel noise are required.
Two examples are scientific charge-coupled devices (CCDs) and complementary metal-oxide-semiconductor (CMOS) active-pixel sensors (APS), which have demonstrated the capability of resolving even single charges per pixel~\cite{Tiffenberg:2017aac,Ma:17}.

The silicon chips themselves (\emph{i.e.}, the ``bare die'') can be made highly radiopure because of the high quality of the silicon wafers, the high chemical purity of the microstructures deposited on the substrate, and the high cleanliness standards in the semiconductor industry.
In addition, if the chip is manufactured following a standard fabrication process in industry, the production capabilities are extremely large.
However, 
standard fabrication limits the thickness of the monolithic devices to the thickness of a standard wafer, which is $\sim$1\,mm.
Hybrid devices can be thicker but require an additional fabrication step to couple the target to the readout chip, which introduces risks for radiocontamination both in the material of the target and through the specific process employed.
The main challenge for both monolithic and hybrid devices is the packaging, \emph{i.e.}, making all the connections for the signals to drive and read the device, usually by bonding to a Kapton flexible cable, while maintaining low radioactive backgrounds in a thin device that has limited capability for self shielding.
Steps to reduce backgrounds include packaging the devices in a clean, radon-free environment to mitigate surface contamination ($\emph{e.g.}$, dust, \pbten) and underground to mitigate cosmogenic activation of the target ($\emph{e.g.}$, tritium~\cite{Saldanha:2020ubf}).
Other external backgrounds from the support structure (device holders, etc.) inside the shield can be minimized with the careful selection of standard radiopure materials.

In recent years, the DAMIC dark matter search demonstrated the 
ability
to build a pixelated silicon detector\textemdash an array of seven scientific CCDs (40\,g of silicon)\textemdash deep underground with low background ($\sim$10 \dru\ below 10\,k\eve \footnote{The unit keV electron-equivalent (k\eve ) is a measure of the amplitude of the ionization signal, \emph{i.e.}, the number of free charge carriers generated by a fast electron with initial kinetic energy of 1\,keV that deposits its full energy in the target.}). Construction is underway of DAMIC-M, a much larger array of 208 CCDs (700\,g of silicon) with 100 times lower background ($\sim$0.1 \dru ). DAMIC also demonstrated that its high spatial resolution can be used to classify particles by event topology and to identify radioactive decay sequences by spatial correlations. These techniques can be employed to suppress backgrounds in the Selena Neutrino Experiment, which will consist of an array of 10,000 hybrid amorphous \ise / CMOS devices (10\,ton of \ise ) for the search for neutrinoless \bb\ decay and for \nue\ detection. Preliminary estimates suggest that zero-background spectroscopy of \bb\ decay and solar neutrinos could be performed in a 100\,ton-year exposure of Selena.

\section{\label{sec:damic}DAMIC dark matter search}

\begin{figure}
\includegraphics[width=\textwidth]{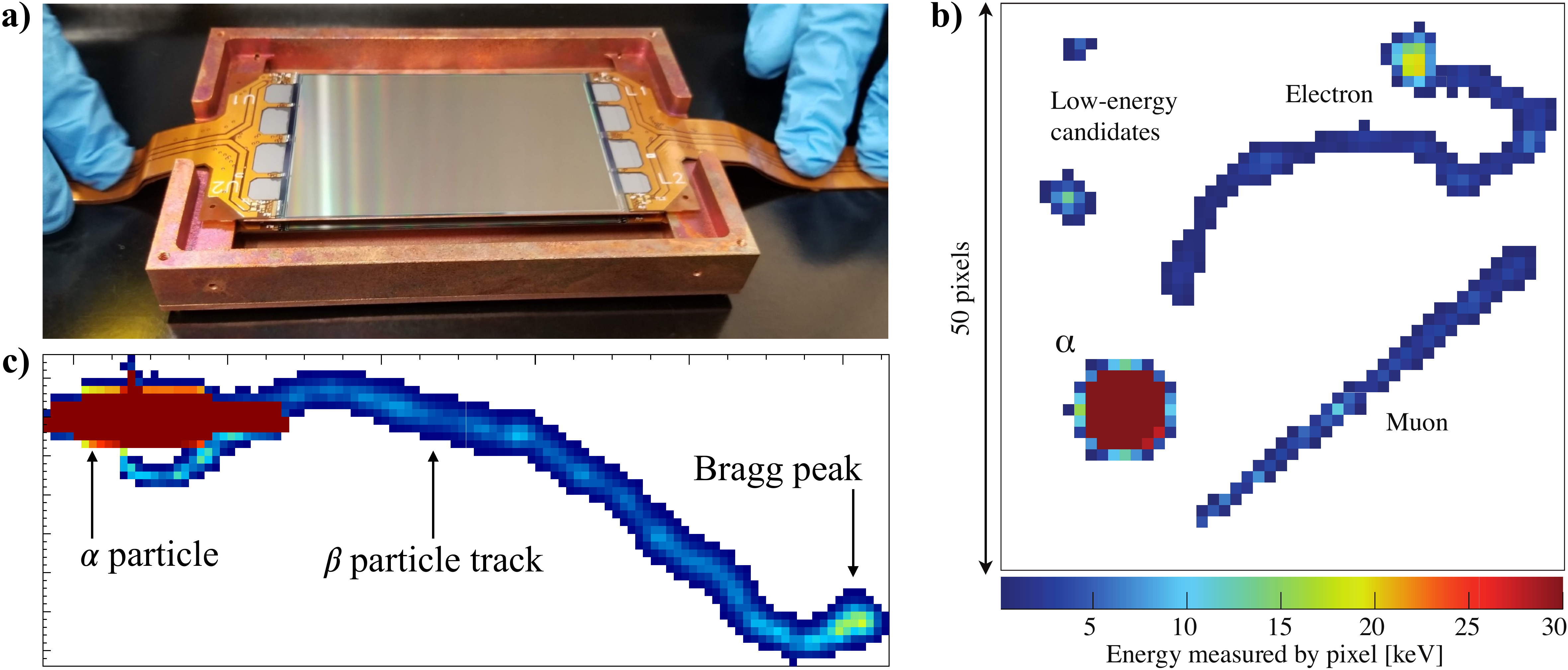}
\caption{\label{fig:damic} {\bf a)}~Photograph of 24-megapixel DAMIC-M CCDs of 9\,cm\,$\times$\,6\,cm in area being installed in a high-purity copper box. The pairs of flex cables that carry the signals to and from the CCDs are glued to the silicon, with wire bonds making the connections to the CCDs. {\bf b)}~A 50$\times$50-pixels segment of a CCD image from an exposure in the surface lab. Different types of particles (labeled) can be easily distinguished by their topology. {\bf c)} Identification of a decay sequence: an $\alpha$ and a $\beta$ particle that occur within an image exposure and originate from the same spatial location in the CCD. The Bragg peak, which occurs at the end of electron/$\beta$ tracks, can be observed.}
\end{figure}

The DAMIC experiment employs scientific CCDs to search for the low-energy ionization signals (nuclear~\cite{DAMIC:2020cut} or electronic~\cite{DAMIC:2016qck,DAMIC:2019dcn} recoils) that may arise from the interaction of dark matter particles in the Galactic Halo with silicon atoms in the target.
DAMIC CCDs are fully-depleted monolithic devices  675-\um\ in thickness and up to 9\,cm\,$\times$\,6\,cm in area (Fig.~\ref{fig:damic}a), with pixels 15\,\um\,$\times$\,15\,\um\ in size~\cite{1185186}.
Figure~\ref{fig:damic}b shows a segment of an image captured with a DAMIC CCD in the surface laboratory.
Different types of particles are identified and labeled based on the pixel-cluster topology.
DAMIC's dark-matter search focuses on the ``low-energy candidates,'' in the keV range and below, which is where the dark-matter signal is expected.
At these energies, the tracks of both electronic and nuclear recoils are much shorter than the pixel size and the shape of the pixel cluster from an event is dominated by charge diffusion, \emph{i.e.,} the spread of the charge cloud as it is drifted from the interaction point to the pixel array.
Since events further away from the pixel array experience more charge diffusion, the spread of the cluster in the image ($x$-$y$ plane) is used to reconstruct the depth ($z$) of the interaction in the target.
The dark-matter search was performed by comparing the energy spectrum and $(x, y, z)$ location of the observed low-energy candidates to the prediction from the radioactive-background model.
The most recent results are presented in Ref.~\onlinecite{DAMIC:2020cut}, while Ref.~\onlinecite{DAMIC:2021crr} provides all details on the construction of the background model, including the detector description and the extensive radioassay program of all components.
The background model was constrained and validated with several independent measurements of radiocontaminants in the detector, \emph{e.g.}, surface/bulk $^{210}$Pb and bulk $^{32}$Si, that were performed with the CCDs themselves by leveraging 
the background-identification techniques described in the next section.

\subsection{\label{sec:backgdstrat}Particle and decay identification}

The classification of particles based on the cluster topology is described in Ref.~\onlinecite{DAMIC:2020wkw}.
Event-by-event classification is most efficient at high deposited energies ($E\gtrsim 100$\,k\eve ), where the electron tracks become much longer than the pixel size.
Alpha particles from the \ueight\ and \thtwo\ decay chains have initial kinetic energies of $\sim$5\,MeV, but 
an $\alpha$ particle that originates on the surface  
loses a fraction of its energy in the \um-thick inactive surface layer of the CCD
and so
may deposit a smaller energy in the target.
DAMIC reported to be fully efficient in the selection of $\alpha$ particles with $E>500$\,k\eve , with a $<0.1$\,\% leakage from electron tracks.
The $\alpha$ particles that occur on the front surface of a CCD can also be separated from those that occur on the back surface based on their topologies.
This strategy was recently extended to the identification of nuclear recoils from neutron scattering~\cite{Neutroninprep}, which, similarly to $\alpha$ particles, generate small, dense charge clusters.
DAMIC is fully efficient in the selection of nuclear recoils with $E>100$\,k\eve , with a leakage of $0.1$\,\% from electron tracks.
Work is ongoing to extend this analysis toward lower energies.

\begin{figure}
\includegraphics[width=\textwidth]{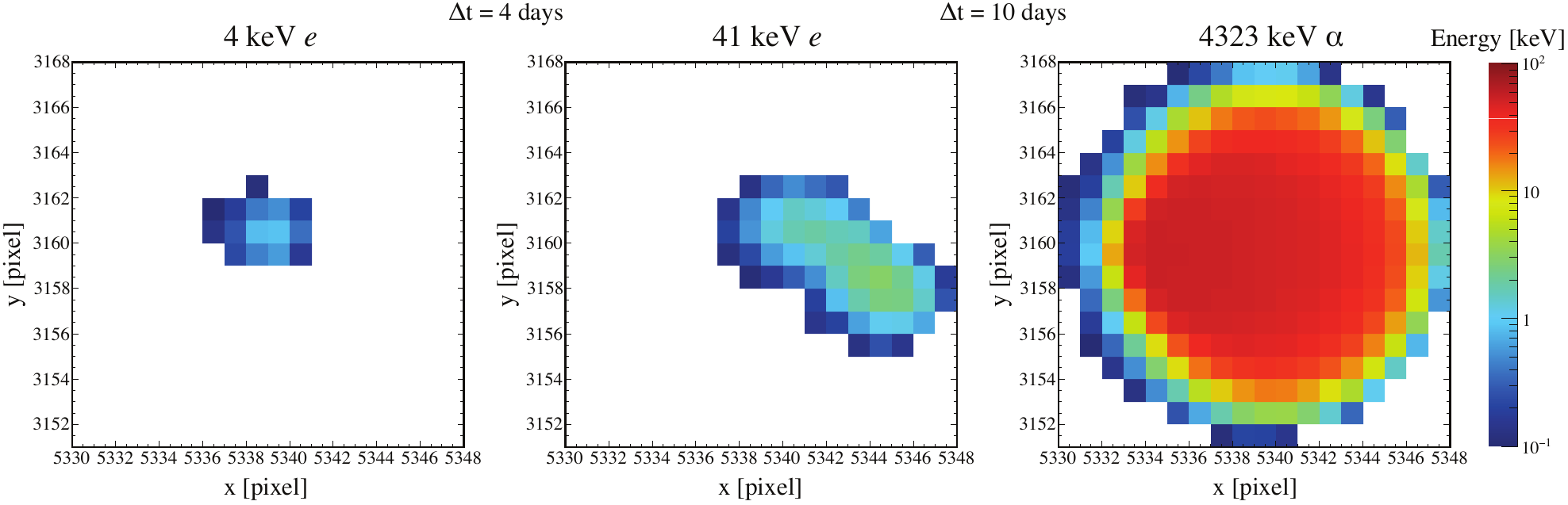}
\caption{\label{fig:evseq}The decay chain of a single \pbten\ nucleus: a sequence of two $\beta$s and one $\alpha$ particle separated in time by several days and detected at the same location.}
\end{figure}

DAMIC can identify radioactive-decay sequences from the spatial correlation between the decay products.
Figure~\ref{fig:damic}c shows an $\alpha$-$\beta$ decay sequence (\emph{e.g.}, $^{214}$BiPo) on the front surface of a CCD, where both decays occur within the same image exposure.
The $\beta$ track starts at the location of the high-energy deposition from the $\alpha$ particle and ends a few mm away where the Bragg peak can be clearly observed.
Decay sequences can also be identified with high efficiency if they occur in separate image exposures.
Fig.~\ref{fig:evseq} shows the identification of the decay of a single \pbten\ nucleus on the CCD surface from a DAMIC run underground, where the three isotopes in the chain decay in separate image exposures many days apart.
The low-energy decay from \pbten\ (Q-value: 63\,keV, $\tau_{1/2}$: 22\,y) spatially coincides with the start-point of the $\beta$ track from \biten\ decay (Q-value: 1.2\,MeV, $\tau_{1/2}$: 5.0\,d) four days later.
At the same spatial location, the $\alpha$ particle from \poten\ decay (Q-value: 5.3\,MeV, $\tau_{1/2}$: 138\,d) was detected 10 days later.
The probability of an accidental $\beta$-$\beta$-$\alpha$ sequence occurring at the same spatial location\textemdash equivalent to $\sim$1\,\textmu g of silicon\textemdash within 14\,days is negligible.

Spatial-coincidence searches were employed to screen DAMIC's silicon target for radiocontaminants, with results reported in Ref.~\onlinecite{DAMIC:2020wkw}.
Remarkably, upper limits at the level of $\sim$10\,\textmu Bq/kg were placed on almost every isotope in the \ueight\ and \thtwo\ decay chains, except for the $\alpha$ emitter $^{230}$Th in the \ueight\ chain, whose decay is separated in time from the previous and subsequent decays by thousands of years.
Of particular relevance was the measurement of the decay rate of the naturally occurring $\beta$ emitter \sitwo\ (Q-value: 230\,keV, $\tau_{1/2}$: 153\,y) from its spatial correlation with the $\beta$ decay of its daughter \ptwo\ (Q-value: 1.7\,MeV, $\tau_{1/2}$: 14\,d).
Since \sitwo\ cannot be removed by chemical purification when the silicon crystal is grown~\cite{Orrell:2017rid}, the \sitwo\ content of the starting material could limit the sensitivity of silicon detectors that search for dark matter~\cite{SuperCDMS:2016wui}.
The measured activity of $140\pm30$\,\textmu Bq/kg has a significantly reduced uncertainty from the first measurement with previous DAMIC CCDs~\cite{DAMIC:2015ipv}, and hints at a variation in \sitwo\ content between different silicon-crystal samples.
For DAMIC-M, the strategy to identify $^{32}$SiP by spatial correlation will allow the experiment to tag and discard this background, with its efficiency limited by the detector's duty cycle.

Recoiling silicon atoms (``nuclear recoils'') following interactions of neutrons or weakly-interacting particles cause atomic dislocations and leave defects in the silicon lattice.
The semiconductor band gap is locally distorted at the location of the defect, which leads to spatially localized spikes in leakage current that are more evident at higher temperatures.
Thus, it may be possible to identify nuclear recoils from the spatial correlation of the ionization cluster with defects that are later identified in images at higher temperature~\cite{Neutroninprep}.
A preliminary analysis of CCD images acquired at 195\,K (compared to the standard operating temperature of 120\,K) after a CCD was irradiated with an AmBe neutron source demonstrates the appearance of dark current spikes spatially correlated with nuclear recoils as identified by cluster topology.
A detailed analysis is ongoing.

\section{\label{sec:selena}Selena Neutrino Experiment}

Selena will be a neutrino observatory with a scientific program that spans fundamental neutrino physics and solar astrophysics. The detector will consist of towers of imaging modules (10,000 modules in total) made from $\sim$5-mm-thick amorphous selenium (aSe)\textemdash isotopically enriched in \ise \textemdash deposited on 300-mm-diameter silicon wafers instrumented with CMOS APS.  The experiment aims to search for the neutrinoless $\beta\beta$ decay of $^{82}$Se, and to perform solar-neutrino spectroscopy and sterile-neutrino searches based on $\nu_e$ capture on $^{82}$Se. Ref.~\onlinecite{Chavarria:2022hwx} presents the scientific motivation for Selena and the conceptual design of a 10-ton detector. 

Selena will leverage 
the event-classification capabilities demonstrated by DAMIC to achieve unprecedentedly low backgrounds. 
This strategy is particularly efficient in its use of the isotopically-enriched active target since background suppression does not rely on self-shielding.
The background estimates presented in the next sections assume raw background rates comparable to those already achieved by kg-scale solid-state detectors for rare event searches, \emph{e.g.}, the M{\footnotesize AJORANA} D{\footnotesize EMON\-STRAT\-OR}~\cite{Majorana:2019nbd}. 
Thus, the requirements in the radiopurity of the construction materials and processes, although challenging, are already possible.
Finally, Selena will operate at room temperature, the CMOS APS for its imaging modules will be fabricated with standard commercial CMOS foundry processes, and the aSe will be deposited following industrial-scale processes developed for the fabrication of medical devices, making the project cost-effective at scale.

Beyond background suppression, the exquisite imaging capabilities of Selena would also enable the study of the neutrinoless \bb\ decay mechanism from the angular correlations between the outgoing electrons~\cite{Ali:2007ec, Deppisch:2020mxv}.
Thus, in the fortunate event that one of the upcoming ton-scale experiments~\cite{Agostini:2021kba} discovers neutrinoless \bb\ decay, Selena would be an ideal follow-up experiment.

\begin{figure}
\includegraphics[width=\textwidth]{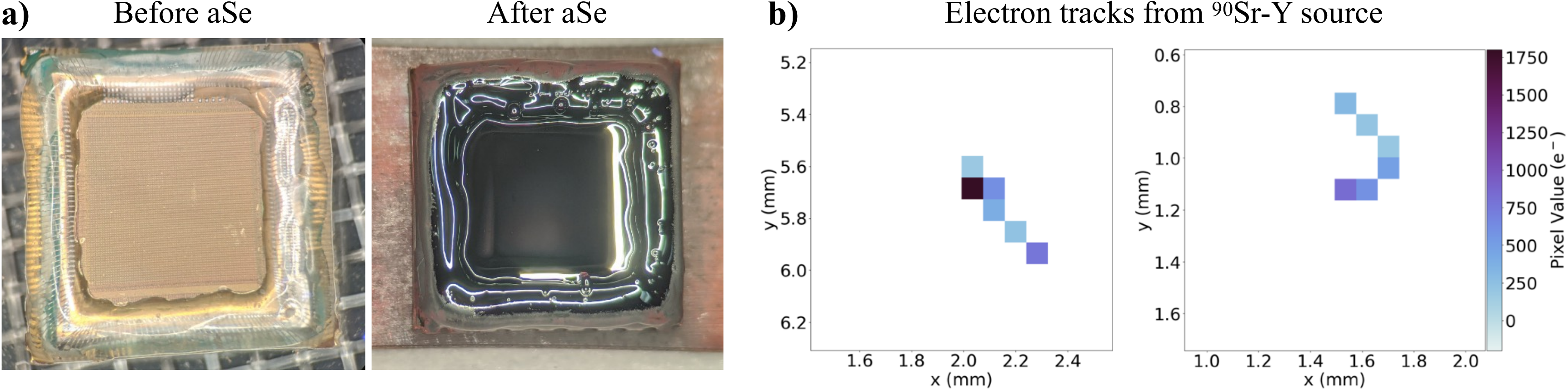}
\caption{\label{fig:selena}{\bf a)} \emph{Topmetal-II$^{\mbox{-}}$} CMOS APS before and after aSe deposition. {\bf b)} Electron tracks observed with the first hybrid aSe/CMOS prototype sensor operated at 4\,V/\um\ at room temperature.}
\end{figure}

The Selena program is currently in its R\&D stage.
The immediate goal is to develop a hybrid aSe/CMOS device that can image electron tracks in the aSe target with comparable resolution to a silicon CCD.
However, the pixel noise requirements are less stringent than DAMIC since the nuclear processes studied by Selena are at higher energies ($\sim$MeV).
Figure~\ref{fig:selena}a shows a recent R\&D device that achieved the lowest pixel noise ever in an aSe imager of 23\,$e^-$ RMS.
The device is based on the \emph{Topmetal-II$^{\mbox{-}}$} CMOS APS from Berkeley Lab~\cite{An:2015oba}, which features a rectangular array of 72$\times$72 pixels with 83-\textmu m pitch.
A 500-\textmu m-thick layer of aSe was deposited by Hologic Corporation on top of the bare metallic contacts of the chip, followed by a thin gold contact to apply the high voltage across the aSe.
The device demonstrated the capability to drift charge in the aSe and collect it on the CMOS APS.
Figure~\ref{fig:selena}b shows the electron tracks imaged by illuminating the device with $\beta$s from a $^{90}$Sr-Y source.
This was the first demonstration of single-particle detection in a hybrid aSe/CMOS pixelated device.
Development is ongoing of a new CMOP APS\textemdash  \emph{Topmetal-Se}\textemdash with 15-\um\ pixel pitch and optimized for charge sensing in aSe.

\subsection{\label{sec:bbbackgd}Backgrounds for neutrinoless \bb\ decay search}

\begin{figure}
\includegraphics[width=\textwidth]{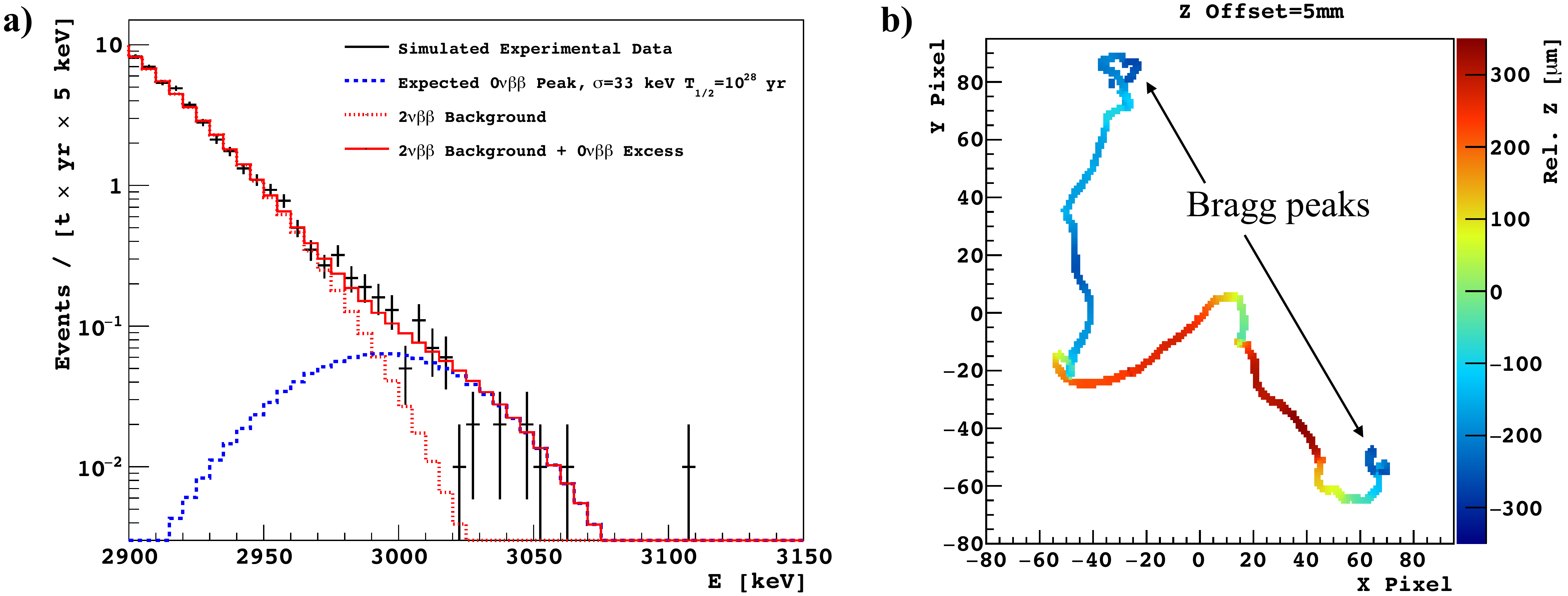}
\caption{\label{fig:bb}{\bf a)} Predicted spectrum from \bb\ decay of $^{82}$Se about $Q_{\beta\beta}$ in Selena, with half-life 
for the neutrinoless (two-neutrino) channel of $\tau_{0\nu}=10^{28}$\,y ($\tau_{2\nu}=10^{20}$\,y).
{\bf b)} Simulated \bb\ decay event with 3.0\,MeV in Selena. In this realization, the CMOS APS has the functionality to measure the time-of-arrival of the ionization charge, which provides relative depth ($z$) reconstruction (color axis). Absolute $z$ offset could be reconstructed from the charge diffusion along the tack. The two Bragg peaks (one for each electron) are clearly visible.}
\end{figure}

Figure~\ref{fig:bb}a shows the expected energy spectrum of \bb\ decays in \ise\ about $Q_{\beta\beta}=3$\,MeV, assuming a half-life of the neutrinoless decay channel $\tau_{0\nu}=10^{28}$\,y.
The energy resolution of aSe was obtained from an ionization-response model validated with 122-keV $\gamma$ rays~\cite{Li:2020ryk}.
The original background study for Selena~\cite{Chavarria:2016hxk} estimates the expected background for \bb\ decay spectroscopy in the region of interest (ROI) about $Q_{\beta\beta}$ to be $6\times10^{-5}$ per keV$_{\rm ee}$$\cdot$ton$\cdot$year.
This extremely low background is possible because of a combination of factors.
First, the high $Q_{\beta\beta}$ of \ise\ is at an energy greater than most backgrounds from primordial $^{238}$U and $^{232}$Th radiocontaminants, which leads to a relatively low event rate.
Second, spatio-temporal correlations effectively reject 
nearly all radioactive decays in the bulk or on the surfaces of the imaging modules.
Finally, external $\gamma$-ray backgrounds, which mostly produce single-electron events from Compton scattering or photoelectric absorption, are suppressed by the requirement that the \bb\ signal events have two clearly identified Bragg peaks (Fig.~\ref{fig:bb}b). This selection retains 50\% of signal events while rejecting 99.9\% of the single-electron background~\cite{Chavarria:2016hxk}.

As an example, consider $^{214}$Bi from the \ueight\ decay chain, a particularly dangerous radioisotope for neutrinoless \bb\ decay searches since its Q-value (3.27\,MeV) is above $Q_{\beta\beta}$ for most candidate isotopes.
None of the $\gamma$ rays emitted in the decay of $^{214}$Bi have sufficient energy to produce a Compton-scattered electron with energy $>2.95$\,MeV.
Thus, the dominant background from $^{214}$Bi in Selena comes from the photoelectric absorption of the 3.05\,MeV $\gamma$ ray (0.02\% intensity).
As mentioned above, the resulting photoelectron can be rejected with high efficiency by event topology.
Thus, the probability that a decay from $^{214}$Bi outside the aSe target produces a background in Selena's ROI is at most (in the case all $\gamma$ rays have their first interaction in the active target) $10^{-9}$!
An analogous calculation results in a similar probability for a background event from $^{208}$Tl in the \thtwo\ decay chain.

The background estimate in Ref.~\onlinecite{Chavarria:2016hxk} also includes backgrounds from higher-energy $\gamma$ rays, \emph{e.g.}, from cosmogenic isotopes activated in-situ or neutron captures in detector components.
Although their flux is generally much lower than $\gamma$ rays from the \ueight\ and \thtwo\ chains, Selena is relatively susceptible to them since they can more easily mimic \bb\ decay.
For example, a high-energy $\gamma$ ray can produce a double-electron event by pair-production in the aSe with the subsequent annihilation $\gamma$ rays escaping the detector module.
Ref.~\onlinecite{Chavarria:2016hxk} argues that this background can be brought to a level below that from \ueight\ and \thtwo\ $\gamma$ rays by minimizing the dead material between modules and by careful selection and procurement of detector components.

\subsection{\label{sec:neubackgd}Backgrounds for solar \nue\ spectroscopy}

\begin{figure}
\includegraphics[width=\textwidth]{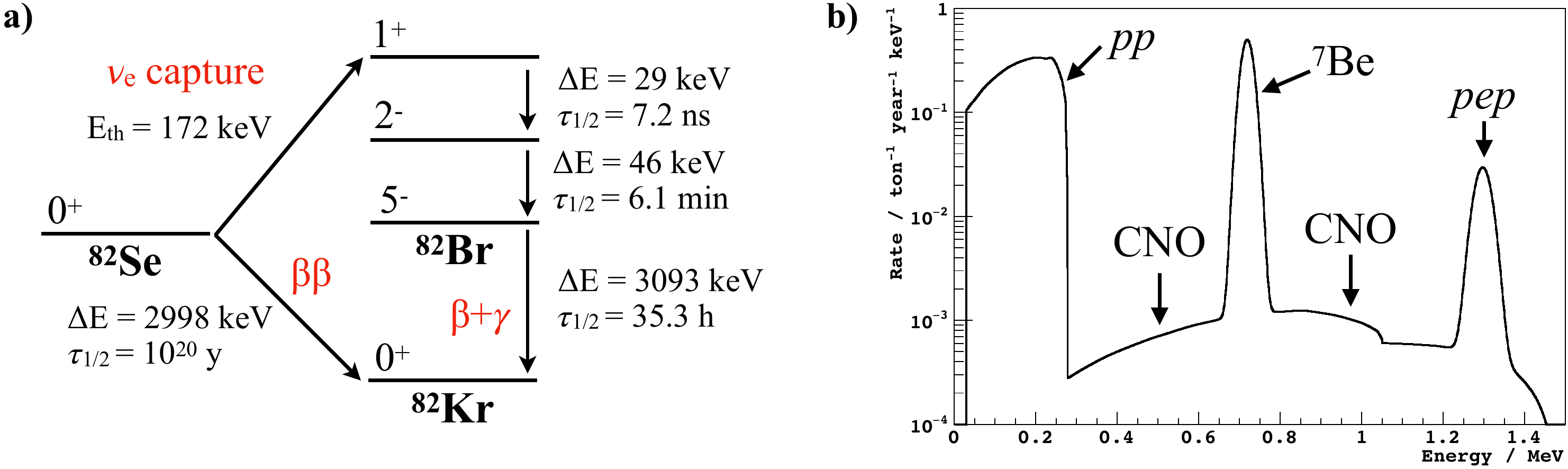}
\caption{\label{fig:solarnu}{\bf a)} Diagram showing two natural channels for $^{82}$Se to transmute into $^{82}$Kr. Both $\beta\beta$ decay and solar-$\nu_e$ capture will be detected with high efficiency by Selena for nuclear-physics studies. {\bf b)} Spectrum of the prompt event from solar-$\nu_e$ capture with energy resolution from the ionization-response model from Ref.~\onlinecite{Li:2020ryk}. The neutrino species from different solar-fusion reactions are labeled. Capture rates from Ref.~\onlinecite{Frekers:2016haa}.}
\end{figure}

Electron neutrinos (\nue) of energy $E_\nu>172$\,keV can capture in \ise\ and lead to the decay sequence shown in Figure~\ref{fig:solarnu}a.
The first step with $\tau_{1/2}=7.2$\,ns is too fast to be distinguished from the electron of energy $E_\nu-172$\,keV emitted following $\nu_e$ capture and together constitute the ``prompt'' event.
The prompt event is followed by a sequence of two decays ($^{82m}$Br$\to$$^{82}$Br$\to$$^{82}$Kr), which occur at the same spatial location with time delays of $\sim$5\,min and $\sim$1\,day, respectively.
By the identification of this ``triple sequence,'' Selena can tag \nue\ captures with efficiency that is only limited by the detector's duty cycle, and without interfering with the search for neutrinoless $\beta\beta$ decay.
The low threshold of the capture reaction provides Selena with sensitivity to all solar neutrino species, with a total of 8500 solar \nue\ captures in a 100-ton-year exposure.
The spectrum of the prompt events, which carry the information of the \nue\ energy, is shown in Figure~\ref{fig:solarnu}b.

The number of accidentals that mimic the triple sequence in a 100 ton-year exposure of an underground detector\textemdash whose background rate is dominated by two-neutrino \bb\ decay\textemdash is $<$$10^{-4}$.
Charged-particle reactions to produce $^{82m}$Br from \ise \textemdash the traditional background for radiochemical experiments\textemdash have a prompt event topology that is easily distinguishable from a fast electron.
A more likely background for the production of $^{82m}$Br would be from neutron capture on trace $^{81}$Br in the aSe, where the deexcitation $\gamma$ ray only interacts once very close to the decay site to mimic the prompt event.
However, the trace contamination of Br in aSe is not known at this time.
The dominant background will likely come from neutron captures on $^{82,80,78}$Se, which lead to sequences of three or four decays that emit fast electrons and $\gamma$ rays.
However, they can all be suppressed by appropriate selections on event topology, time separation, and decay energies.
Finally, a survey of isotopes that can be cosmogenically activated in $^{82}$Se did not identify any that could be produced at any significant rates and could mimic the triple sequence closely enough in topology and time separation.
Thus, so far, it appears that zero-background solar-neutrino spectroscopy is possible in a 100-ton-year exposure with Selena.

\section{\label{sec:conclusion}Conclusion}
The DAMIC Collaboration demonstrated that the high spatial resolution of low-noise silicon CCDs can be used to classify particles (\emph{e.g.}, $\alpha$, $\beta$/fast electrons, nuclear recoils) by event topology and to identify radioactive decay sequences by spatial correlations, even for time separations between decays of many weeks~\cite{DAMIC:2020wkw}.
These techniques were already applied in the DAMIC experiment to constrain the background model for its recent dark matter search~\cite{DAMIC:2021crr}.
The Selena Neutrino Experiment aims for a background-suppression strategy based on these techniques to perform \bb\ decay and solar \nue\ spectroscopy with zero background in a 100\,ton-year exposure of \ise ~\cite{Chavarria:2022hwx}.
For this purpose, development is ongoing of novel aSe/CMOS hybrid sensors to read out the ionization charge in an aSe target with high resolution.
The first prototype devices already demonstrated the capability to image fast-electron tracks.

\begin{acknowledgments}
These proceedings are based on years of work by many scientists in the DAMIC and Selena Collaborations.
\end{acknowledgments}

\nocite{*}
\bibliography{myrefs}

\begin{thebibliography}{23}%
\makeatletter
\providecommand \@ifxundefined [1]{%
 \@ifx{#1\undefined}
}%
\providecommand \@ifnum [1]{%
 \ifnum #1\expandafter \@firstoftwo
 \else \expandafter \@secondoftwo
 \fi
}%
\providecommand \@ifx [1]{%
 \ifx #1\expandafter \@firstoftwo
 \else \expandafter \@secondoftwo
 \fi
}%
\providecommand \natexlab [1]{#1}%
\providecommand \enquote  [1]{``#1''}%
\providecommand \bibnamefont  [1]{#1}%
\providecommand \bibfnamefont [1]{#1}%
\providecommand \citenamefont [1]{#1}%
\providecommand \href@noop [0]{\@secondoftwo}%
\providecommand \href [0]{\begingroup \@sanitize@url \@href}%
\providecommand \@href[1]{\@@startlink{#1}\@@href}%
\providecommand \@@href[1]{\endgroup#1\@@endlink}%
\providecommand \@sanitize@url [0]{\catcode `\\12\catcode `\$12\catcode
  `\&12\catcode `\#12\catcode `\^12\catcode `\_12\catcode `\%12\relax}%
\providecommand \@@startlink[1]{}%
\providecommand \@@endlink[0]{}%
\providecommand \url  [0]{\begingroup\@sanitize@url \@url }%
\providecommand \@url [1]{\endgroup\@href {#1}{\urlprefix }}%
\providecommand \urlprefix  [0]{URL }%
\providecommand \Eprint [0]{\href }%
\providecommand \doibase [0]{http://dx.doi.org/}%
\providecommand \selectlanguage [0]{\@gobble}%
\providecommand \bibinfo  [0]{\@secondoftwo}%
\providecommand \bibfield  [0]{\@secondoftwo}%
\providecommand \translation [1]{[#1]}%
\providecommand \BibitemOpen [0]{}%
\providecommand \bibitemStop [0]{}%
\providecommand \bibitemNoStop [0]{.\EOS\space}%
\providecommand \EOS [0]{\spacefactor3000\relax}%
\providecommand \BibitemShut  [1]{\csname bibitem#1\endcsname}%
\let\auto@bib@innerbib\@empty
\bibitem [{\citenamefont {Tiffenberg}\ \emph {et~al.}(2017)\citenamefont
  {Tiffenberg}, \citenamefont {Sofo-Haro}, \citenamefont {Drlica-Wagner},
  \citenamefont {Essig}, \citenamefont {Guardincerri}, \citenamefont {Holland},
  \citenamefont {Volansky},\ and\ \citenamefont {Yu}}]{Tiffenberg:2017aac}%
  \BibitemOpen
  \bibfield  {author} {\bibinfo {author} {\bibfnamefont {J.}~\bibnamefont
  {Tiffenberg}}, \bibinfo {author} {\bibfnamefont {M.}~\bibnamefont
  {Sofo-Haro}}, \bibinfo {author} {\bibfnamefont {A.}~\bibnamefont
  {Drlica-Wagner}}, \bibinfo {author} {\bibfnamefont {R.}~\bibnamefont
  {Essig}}, \bibinfo {author} {\bibfnamefont {Y.}~\bibnamefont {Guardincerri}},
  \bibinfo {author} {\bibfnamefont {S.}~\bibnamefont {Holland}}, \bibinfo
  {author} {\bibfnamefont {T.}~\bibnamefont {Volansky}}, \ and\ \bibinfo
  {author} {\bibfnamefont {T.-T.}\ \bibnamefont {Yu}} (\bibinfo {collaboration}
  {SENSEI Collaboration}),\ }\bibfield  {title} {\enquote {\bibinfo {title}
  {{Single-electron and single-photon sensitivity with a silicon Skipper
  CCD}},}\ }\href {\doibase 10.1103/PhysRevLett.119.131802} {\bibfield
  {journal} {\bibinfo  {journal} {Phys. Rev. Lett.}\ }\textbf {\bibinfo
  {volume} {119}},\ \bibinfo {pages} {131802} (\bibinfo {year} {2017})},\
  \Eprint {http://arxiv.org/abs/1706.00028} {arXiv:1706.00028
  [physics.ins-det]} \BibitemShut {NoStop}%
\bibitem [{\citenamefont {Ma}\ \emph {et~al.}(2017)\citenamefont {Ma},
  \citenamefont {Masoodian}, \citenamefont {Starkey},\ and\ \citenamefont
  {Fossum}}]{Ma:17}%
  \BibitemOpen
  \bibfield  {author} {\bibinfo {author} {\bibfnamefont {J.}~\bibnamefont
  {Ma}}, \bibinfo {author} {\bibfnamefont {S.}~\bibnamefont {Masoodian}},
  \bibinfo {author} {\bibfnamefont {D.~A.}\ \bibnamefont {Starkey}}, \ and\
  \bibinfo {author} {\bibfnamefont {E.~R.}\ \bibnamefont {Fossum}},\ }\bibfield
   {title} {\enquote {\bibinfo {title} {Photon-number-resolving megapixel image
  sensor at room temperature without avalanche gain},}\ }\href {\doibase
  10.1364/OPTICA.4.001474} {\bibfield  {journal} {\bibinfo  {journal} {Optica}\
  }\textbf {\bibinfo {volume} {4}},\ \bibinfo {pages} {1474--1481} (\bibinfo
  {year} {2017})}\BibitemShut {NoStop}%
\bibitem [{\citenamefont {Saldanha}\ \emph {et~al.}(2020)\citenamefont
  {Saldanha}, \citenamefont {Thomas}, \citenamefont {Tsang}, \citenamefont
  {Chavarria}, \citenamefont {Bunker}, \citenamefont {Burnett}, \citenamefont
  {Elliott}, \citenamefont {Matalon}, \citenamefont {Mitra}, \citenamefont
  {Piers}, \citenamefont {Privitera}, \citenamefont {Ramanathan},\ and\
  \citenamefont {Smida}}]{Saldanha:2020ubf}%
  \BibitemOpen
  \bibfield  {author} {\bibinfo {author} {\bibfnamefont {R.}~\bibnamefont
  {Saldanha}}, \bibinfo {author} {\bibfnamefont {R.}~\bibnamefont {Thomas}},
  \bibinfo {author} {\bibfnamefont {R.~H.~M.}\ \bibnamefont {Tsang}}, \bibinfo
  {author} {\bibfnamefont {A.~E.}\ \bibnamefont {Chavarria}}, \bibinfo {author}
  {\bibfnamefont {R.}~\bibnamefont {Bunker}}, \bibinfo {author} {\bibfnamefont
  {J.~L.}\ \bibnamefont {Burnett}}, \bibinfo {author} {\bibfnamefont {S.~R.}\
  \bibnamefont {Elliott}}, \bibinfo {author} {\bibfnamefont {A.}~\bibnamefont
  {Matalon}}, \bibinfo {author} {\bibfnamefont {P.}~\bibnamefont {Mitra}},
  \bibinfo {author} {\bibfnamefont {A.}~\bibnamefont {Piers}}, \bibinfo
  {author} {\bibfnamefont {P.}~\bibnamefont {Privitera}}, \bibinfo {author}
  {\bibfnamefont {K.}~\bibnamefont {Ramanathan}}, \ and\ \bibinfo {author}
  {\bibfnamefont {R.}~\bibnamefont {Smida}},\ }\bibfield  {title} {\enquote
  {\bibinfo {title} {{Cosmogenic activation of silicon}},}\ }\href {\doibase
  10.1103/PhysRevD.102.102006} {\bibfield  {journal} {\bibinfo  {journal}
  {Phys. Rev. D}\ }\textbf {\bibinfo {volume} {102}},\ \bibinfo {pages}
  {102006} (\bibinfo {year} {2020})},\ \Eprint
  {http://arxiv.org/abs/2007.10584} {arXiv:2007.10584 [physics.ins-det]}
  \BibitemShut {NoStop}%
\bibitem [{Note1()}]{Note1}%
  \BibitemOpen
  \bibinfo {note} {The unit keV electron-equivalent (k\protect \mbox
  {eV$_{\protect \rm ee}$}) is a measure of the amplitude of the ionization
  signal, \protect \emph {i.e.}, the number of free charge carriers generated
  by a fast electron with initial kinetic energy of 1\protect \,keV that
  deposits its full energy in the target.}\BibitemShut {Stop}%
\bibitem [{\citenamefont {Aguilar-Arevalo}\ \emph {et~al.}(2020)\citenamefont
  {Aguilar-Arevalo} \emph {et~al.}}]{DAMIC:2020cut}%
  \BibitemOpen
  \bibfield  {author} {\bibinfo {author} {\bibfnamefont {A.}~\bibnamefont
  {Aguilar-Arevalo}} \emph {et~al.} (\bibinfo {collaboration} {DAMIC
  Collaboration}),\ }\bibfield  {title} {\enquote {\bibinfo {title} {{Results
  on low-mass weakly interacting massive particles from a 11 kg-day target
  exposure of DAMIC at SNOLAB}},}\ }\href {\doibase
  10.1103/PhysRevLett.125.241803} {\bibfield  {journal} {\bibinfo  {journal}
  {Phys. Rev. Lett.}\ }\textbf {\bibinfo {volume} {125}},\ \bibinfo {pages}
  {241803} (\bibinfo {year} {2020})},\ \Eprint
  {http://arxiv.org/abs/2007.15622} {arXiv:2007.15622 [astro-ph.CO]}
  \BibitemShut {NoStop}%
\bibitem [{\citenamefont {Aguilar-Arevalo}\ \emph {et~al.}(2017)\citenamefont
  {Aguilar-Arevalo} \emph {et~al.}}]{DAMIC:2016qck}%
  \BibitemOpen
  \bibfield  {author} {\bibinfo {author} {\bibfnamefont {A.}~\bibnamefont
  {Aguilar-Arevalo}} \emph {et~al.} (\bibinfo {collaboration} {DAMIC
  Collaboration}),\ }\bibfield  {title} {\enquote {\bibinfo {title} {{First
  direct-detection constraints on eV-scale hidden-photon dark matter with DAMIC
  at SNOLAB}},}\ }\href {\doibase 10.1103/PhysRevLett.118.141803} {\bibfield
  {journal} {\bibinfo  {journal} {Phys. Rev. Lett.}\ }\textbf {\bibinfo
  {volume} {118}},\ \bibinfo {pages} {141803} (\bibinfo {year} {2017})},\
  \Eprint {http://arxiv.org/abs/1611.03066} {arXiv:1611.03066 [astro-ph.CO]}
  \BibitemShut {NoStop}%
\bibitem [{\citenamefont {Aguilar-Arevalo}\ \emph {et~al.}(2019)\citenamefont
  {Aguilar-Arevalo} \emph {et~al.}}]{DAMIC:2019dcn}%
  \BibitemOpen
  \bibfield  {author} {\bibinfo {author} {\bibfnamefont {A.}~\bibnamefont
  {Aguilar-Arevalo}} \emph {et~al.} (\bibinfo {collaboration} {DAMIC
  Colaboration}),\ }\bibfield  {title} {\enquote {\bibinfo {title}
  {{Constraints on light dark matter particles interacting with electrons from
  DAMIC at SNOLAB}},}\ }\href {\doibase 10.1103/PhysRevLett.123.181802}
  {\bibfield  {journal} {\bibinfo  {journal} {Phys. Rev. Lett.}\ }\textbf
  {\bibinfo {volume} {123}},\ \bibinfo {pages} {181802} (\bibinfo {year}
  {2019})},\ \Eprint {http://arxiv.org/abs/1907.12628} {arXiv:1907.12628
  [astro-ph.CO]} \BibitemShut {NoStop}%
\bibitem [{\citenamefont {Holland}\ \emph {et~al.}(2003)\citenamefont
  {Holland}, \citenamefont {Groom}, \citenamefont {Palaio}, \citenamefont
  {Stover},\ and\ \citenamefont {Wei}}]{1185186}%
  \BibitemOpen
  \bibfield  {author} {\bibinfo {author} {\bibfnamefont {S.}~\bibnamefont
  {Holland}}, \bibinfo {author} {\bibfnamefont {D.}~\bibnamefont {Groom}},
  \bibinfo {author} {\bibfnamefont {N.}~\bibnamefont {Palaio}}, \bibinfo
  {author} {\bibfnamefont {R.}~\bibnamefont {Stover}}, \ and\ \bibinfo {author}
  {\bibfnamefont {M.}~\bibnamefont {Wei}},\ }\bibfield  {title} {\enquote
  {\bibinfo {title} {Fully depleted, back-illuminated charge-coupled devices
  fabricated on high-resistivity silicon},}\ }\href {\doibase
  10.1109/TED.2002.806476} {\bibfield  {journal} {\bibinfo  {journal} {IEEE
  Trans. Electron Devices}\ }\textbf {\bibinfo {volume} {50}},\ \bibinfo
  {pages} {225--238} (\bibinfo {year} {2003})}\BibitemShut {NoStop}%
\bibitem [{\citenamefont {Aguilar-Arevalo}\ \emph {et~al.}(2022)\citenamefont
  {Aguilar-Arevalo} \emph {et~al.}}]{DAMIC:2021crr}%
  \BibitemOpen
  \bibfield  {author} {\bibinfo {author} {\bibfnamefont {A.}~\bibnamefont
  {Aguilar-Arevalo}} \emph {et~al.} (\bibinfo {collaboration} {DAMIC
  Collaboration}),\ }\bibfield  {title} {\enquote {\bibinfo {title}
  {{Characterization of the background spectrum in DAMIC at SNOLAB}},}\ }\href
  {\doibase 10.1103/PhysRevD.105.062003} {\bibfield  {journal} {\bibinfo
  {journal} {Phys. Rev. D}\ }\textbf {\bibinfo {volume} {105}},\ \bibinfo
  {pages} {062003} (\bibinfo {year} {2022})},\ \Eprint
  {http://arxiv.org/abs/2110.13133} {arXiv:2110.13133 [hep-ex]} \BibitemShut
  {NoStop}%
\bibitem [{\citenamefont {Aguilar-Arevalo}\ \emph {et~al.}(2021)\citenamefont
  {Aguilar-Arevalo} \emph {et~al.}}]{DAMIC:2020wkw}%
  \BibitemOpen
  \bibfield  {author} {\bibinfo {author} {\bibfnamefont {A.}~\bibnamefont
  {Aguilar-Arevalo}} \emph {et~al.} (\bibinfo {collaboration} {DAMIC
  Collaboration}),\ }\bibfield  {title} {\enquote {\bibinfo {title}
  {{Measurement of the bulk radioactive contamination of detector-grade silicon
  with DAMIC at SNOLAB}},}\ }\href {\doibase 10.1088/1748-0221/16/06/P06019}
  {\bibfield  {journal} {\bibinfo  {journal} {JINST}\ }\textbf {\bibinfo
  {volume} {16}},\ \bibinfo {pages} {P06019} (\bibinfo {year} {2021})},\
  \Eprint {http://arxiv.org/abs/2011.12922} {arXiv:2011.12922
  [physics.ins-det]} \BibitemShut {NoStop}%
\bibitem [{\citenamefont {Chavarria}\ \emph {et~al.}()\citenamefont {Chavarria}
  \emph {et~al.}}]{Neutroninprep}%
  \BibitemOpen
  \bibfield  {author} {\bibinfo {author} {\bibfnamefont {A.~E.}\ \bibnamefont
  {Chavarria}} \emph {et~al.},\ }\href@noop {} {}\bibinfo {howpublished} {in
  preparation}\BibitemShut {NoStop}%
\bibitem [{\citenamefont {Orrell}\ \emph {et~al.}(2018)\citenamefont {Orrell},
  \citenamefont {Arnquist}, \citenamefont {Bliss}, \citenamefont {Bunker},\
  and\ \citenamefont {Finch}}]{Orrell:2017rid}%
  \BibitemOpen
  \bibfield  {author} {\bibinfo {author} {\bibfnamefont {J.~L.}\ \bibnamefont
  {Orrell}}, \bibinfo {author} {\bibfnamefont {I.~J.}\ \bibnamefont
  {Arnquist}}, \bibinfo {author} {\bibfnamefont {M.}~\bibnamefont {Bliss}},
  \bibinfo {author} {\bibfnamefont {R.}~\bibnamefont {Bunker}}, \ and\ \bibinfo
  {author} {\bibfnamefont {Z.~S.}\ \bibnamefont {Finch}},\ }\bibfield  {title}
  {\enquote {\bibinfo {title} {{Naturally occurring $^{32}$Si and
  low-background silicon dark matter detectors}},}\ }\href {\doibase
  10.1016/j.astropartphys.2018.02.005} {\bibfield  {journal} {\bibinfo
  {journal} {Astropart. Phys.}\ }\textbf {\bibinfo {volume} {99}},\ \bibinfo
  {pages} {9--20} (\bibinfo {year} {2018})},\ \Eprint
  {http://arxiv.org/abs/1708.00110} {arXiv:1708.00110 [physics.ins-det]}
  \BibitemShut {NoStop}%
\bibitem [{\citenamefont {Agnese}\ \emph {et~al.}(2017)\citenamefont {Agnese}
  \emph {et~al.}}]{SuperCDMS:2016wui}%
  \BibitemOpen
  \bibfield  {author} {\bibinfo {author} {\bibfnamefont {R.}~\bibnamefont
  {Agnese}} \emph {et~al.} (\bibinfo {collaboration} {SuperCDMS
  Collaboration}),\ }\bibfield  {title} {\enquote {\bibinfo {title} {{Projected
  sensitivity of the SuperCDMS SNOLAB experiment}},}\ }\href {\doibase
  10.1103/PhysRevD.95.082002} {\bibfield  {journal} {\bibinfo  {journal} {Phys.
  Rev. D}\ }\textbf {\bibinfo {volume} {95}},\ \bibinfo {pages} {082002}
  (\bibinfo {year} {2017})},\ \Eprint {http://arxiv.org/abs/1610.00006}
  {arXiv:1610.00006 [physics.ins-det]} \BibitemShut {NoStop}%
\bibitem [{\citenamefont {Aguilar-Arevalo}\ \emph {et~al.}(2015)\citenamefont
  {Aguilar-Arevalo} \emph {et~al.}}]{DAMIC:2015ipv}%
  \BibitemOpen
  \bibfield  {author} {\bibinfo {author} {\bibfnamefont {A.}~\bibnamefont
  {Aguilar-Arevalo}} \emph {et~al.} (\bibinfo {collaboration} {DAMIC
  Collaboration}),\ }\bibfield  {title} {\enquote {\bibinfo {title}
  {{Measurement of radioactive contamination in the high-resistivity silicon
  CCDs of the DAMIC experiment}},}\ }\href {\doibase
  10.1088/1748-0221/10/08/P08014} {\bibfield  {journal} {\bibinfo  {journal}
  {JINST}\ }\textbf {\bibinfo {volume} {10}},\ \bibinfo {pages} {P08014}
  (\bibinfo {year} {2015})},\ \Eprint {http://arxiv.org/abs/1506.02562}
  {arXiv:1506.02562 [astro-ph.IM]} \BibitemShut {NoStop}%
\bibitem [{\citenamefont {Chavarria}\ \emph {et~al.}(2022)\citenamefont
  {Chavarria}, \citenamefont {Galbiati}, \citenamefont {Hernandez-Molinero},
  \citenamefont {Ianni}, \citenamefont {Li}, \citenamefont {Mei}, \citenamefont
  {Montanino}, \citenamefont {Ni}, \citenamefont {Garay}, \citenamefont
  {Piers},\ and\ \citenamefont {Wang}}]{Chavarria:2022hwx}%
  \BibitemOpen
  \bibfield  {author} {\bibinfo {author} {\bibfnamefont {A.~E.}\ \bibnamefont
  {Chavarria}}, \bibinfo {author} {\bibfnamefont {C.}~\bibnamefont {Galbiati}},
  \bibinfo {author} {\bibfnamefont {B.}~\bibnamefont {Hernandez-Molinero}},
  \bibinfo {author} {\bibfnamefont {A.}~\bibnamefont {Ianni}}, \bibinfo
  {author} {\bibfnamefont {X.}~\bibnamefont {Li}}, \bibinfo {author}
  {\bibfnamefont {Y.}~\bibnamefont {Mei}}, \bibinfo {author} {\bibfnamefont
  {D.}~\bibnamefont {Montanino}}, \bibinfo {author} {\bibfnamefont
  {X.}~\bibnamefont {Ni}}, \bibinfo {author} {\bibfnamefont {C.~P.}\
  \bibnamefont {Garay}}, \bibinfo {author} {\bibfnamefont {A.}~\bibnamefont
  {Piers}}, \ and\ \bibinfo {author} {\bibfnamefont {H.}~\bibnamefont {Wang}},\
  }\bibfield  {title} {\enquote {\bibinfo {title} {{Snowmass 2021 white paper:
  the Selena neutrino experiment}},}\ }in\ \href@noop {} {\emph {\bibinfo
  {booktitle} {{2022 Snowmass Summer Study}}}}\ (\bibinfo {year} {2022})\
  \Eprint {http://arxiv.org/abs/2203.08779} {arXiv:2203.08779
  [physics.ins-det]} \BibitemShut {NoStop}%
\bibitem [{\citenamefont {Alvis}\ \emph {et~al.}(2019)\citenamefont {Alvis}
  \emph {et~al.}}]{Majorana:2019nbd}%
  \BibitemOpen
  \bibfield  {author} {\bibinfo {author} {\bibfnamefont {S.~I.}\ \bibnamefont
  {Alvis}} \emph {et~al.} (\bibinfo {collaboration} {Majorana Collaboration}),\
  }\bibfield  {title} {\enquote {\bibinfo {title} {{A search for neutrinoless
  double-beta decay in $^{76}$Ge with 26 kg-yr of exposure from the MAJORANA
  DEMONSTRATOR}},}\ }\href {\doibase 10.1103/PhysRevC.100.025501} {\bibfield
  {journal} {\bibinfo  {journal} {Phys. Rev. C}\ }\textbf {\bibinfo {volume}
  {100}},\ \bibinfo {pages} {025501} (\bibinfo {year} {2019})},\ \Eprint
  {http://arxiv.org/abs/1902.02299} {arXiv:1902.02299 [nucl-ex]} \BibitemShut
  {NoStop}%
\bibitem [{\citenamefont {Ali}, \citenamefont {Borisov},\ and\ \citenamefont
  {Zhuridov}(2007)}]{Ali:2007ec}%
  \BibitemOpen
  \bibfield  {author} {\bibinfo {author} {\bibfnamefont {A.}~\bibnamefont
  {Ali}}, \bibinfo {author} {\bibfnamefont {A.~V.}\ \bibnamefont {Borisov}}, \
  and\ \bibinfo {author} {\bibfnamefont {D.~V.}\ \bibnamefont {Zhuridov}},\
  }\bibfield  {title} {\enquote {\bibinfo {title} {{Probing new physics in the
  neutrinoless double beta decay using electron angular correlation}},}\ }\href
  {\doibase 10.1103/PhysRevD.76.093009} {\bibfield  {journal} {\bibinfo
  {journal} {Phys. Rev. D}\ }\textbf {\bibinfo {volume} {76}},\ \bibinfo
  {pages} {093009} (\bibinfo {year} {2007})},\ \Eprint
  {http://arxiv.org/abs/0706.4165} {arXiv:0706.4165 [hep-ph]} \BibitemShut
  {NoStop}%
\bibitem [{\citenamefont {Deppisch}, \citenamefont {Graf},\ and\ \citenamefont
  {\v{S}imkovic}(2020)}]{Deppisch:2020mxv}%
  \BibitemOpen
  \bibfield  {author} {\bibinfo {author} {\bibfnamefont {F.~F.}\ \bibnamefont
  {Deppisch}}, \bibinfo {author} {\bibfnamefont {L.}~\bibnamefont {Graf}}, \
  and\ \bibinfo {author} {\bibfnamefont {F.}~\bibnamefont {\v{S}imkovic}},\
  }\bibfield  {title} {\enquote {\bibinfo {title} {Searching for new physics in
  two-neutrino double beta decay},}\ }\href {\doibase
  10.1103/PhysRevLett.125.171801} {\bibfield  {journal} {\bibinfo  {journal}
  {Phys. Rev. Lett.}\ }\textbf {\bibinfo {volume} {125}},\ \bibinfo {pages}
  {171801} (\bibinfo {year} {2020})},\ \Eprint
  {http://arxiv.org/abs/2003.11836} {arXiv:2003.11836 [hep-ph]} \BibitemShut
  {NoStop}%
\bibitem [{\citenamefont {Agostini}\ \emph {et~al.}(2021)\citenamefont
  {Agostini}, \citenamefont {Benato}, \citenamefont {Detwiler}, \citenamefont
  {Men\'endez},\ and\ \citenamefont {Vissani}}]{Agostini:2021kba}%
  \BibitemOpen
  \bibfield  {author} {\bibinfo {author} {\bibfnamefont {M.}~\bibnamefont
  {Agostini}}, \bibinfo {author} {\bibfnamefont {G.}~\bibnamefont {Benato}},
  \bibinfo {author} {\bibfnamefont {J.~A.}\ \bibnamefont {Detwiler}}, \bibinfo
  {author} {\bibfnamefont {J.}~\bibnamefont {Men\'endez}}, \ and\ \bibinfo
  {author} {\bibfnamefont {F.}~\bibnamefont {Vissani}},\ }\bibfield  {title}
  {\enquote {\bibinfo {title} {{Testing the inverted neutrino mass ordering
  with neutrinoless double-\ensuremath{\beta} decay}},}\ }\href {\doibase
  10.1103/PhysRevC.104.L042501} {\bibfield  {journal} {\bibinfo  {journal}
  {Phys. Rev. C}\ }\textbf {\bibinfo {volume} {104}},\ \bibinfo {pages}
  {L042501} (\bibinfo {year} {2021})},\ \Eprint
  {http://arxiv.org/abs/2107.09104} {arXiv:2107.09104 [hep-ph]} \BibitemShut
  {NoStop}%
\bibitem [{\citenamefont {An}\ \emph {et~al.}(2016)\citenamefont {An},
  \citenamefont {Chen}, \citenamefont {Gao}, \citenamefont {Han}, \citenamefont
  {Ji}, \citenamefont {Li}, \citenamefont {Mei}, \citenamefont {Sun},
  \citenamefont {Sun}, \citenamefont {Wang}, \citenamefont {Xiao},
  \citenamefont {Yang},\ and\ \citenamefont {Zhou}}]{An:2015oba}%
  \BibitemOpen
  \bibfield  {author} {\bibinfo {author} {\bibfnamefont {M.}~\bibnamefont
  {An}}, \bibinfo {author} {\bibfnamefont {C.}~\bibnamefont {Chen}}, \bibinfo
  {author} {\bibfnamefont {C.}~\bibnamefont {Gao}}, \bibinfo {author}
  {\bibfnamefont {M.}~\bibnamefont {Han}}, \bibinfo {author} {\bibfnamefont
  {R.}~\bibnamefont {Ji}}, \bibinfo {author} {\bibfnamefont {X.}~\bibnamefont
  {Li}}, \bibinfo {author} {\bibfnamefont {Y.}~\bibnamefont {Mei}}, \bibinfo
  {author} {\bibfnamefont {Q.}~\bibnamefont {Sun}}, \bibinfo {author}
  {\bibfnamefont {X.}~\bibnamefont {Sun}}, \bibinfo {author} {\bibfnamefont
  {K.}~\bibnamefont {Wang}}, \bibinfo {author} {\bibfnamefont {L.}~\bibnamefont
  {Xiao}}, \bibinfo {author} {\bibfnamefont {P.}~\bibnamefont {Yang}}, \ and\
  \bibinfo {author} {\bibfnamefont {W.}~\bibnamefont {Zhou}},\ }\bibfield
  {title} {\enquote {\bibinfo {title} {{A low-noise CMOS pixel direct charge
  sensor, Topmetal-II-}},}\ }\href {\doibase 10.1016/j.nima.2015.11.153}
  {\bibfield  {journal} {\bibinfo  {journal} {Nucl. Instrum. Meth. A}\ }\textbf
  {\bibinfo {volume} {810}},\ \bibinfo {pages} {144--150} (\bibinfo {year}
  {2016})},\ \Eprint {http://arxiv.org/abs/1509.08611} {arXiv:1509.08611
  [physics.ins-det]} \BibitemShut {NoStop}%
\bibitem [{\citenamefont {Li}\ \emph {et~al.}(2021)\citenamefont {Li},
  \citenamefont {Chavarria}, \citenamefont {Bogdanovich}, \citenamefont
  {Galbiati}, \citenamefont {Piers},\ and\ \citenamefont
  {Polischuk}}]{Li:2020ryk}%
  \BibitemOpen
  \bibfield  {author} {\bibinfo {author} {\bibfnamefont {X.}~\bibnamefont
  {Li}}, \bibinfo {author} {\bibfnamefont {A.~E.}\ \bibnamefont {Chavarria}},
  \bibinfo {author} {\bibfnamefont {S.}~\bibnamefont {Bogdanovich}}, \bibinfo
  {author} {\bibfnamefont {C.}~\bibnamefont {Galbiati}}, \bibinfo {author}
  {\bibfnamefont {A.}~\bibnamefont {Piers}}, \ and\ \bibinfo {author}
  {\bibfnamefont {B.}~\bibnamefont {Polischuk}},\ }\bibfield  {title} {\enquote
  {\bibinfo {title} {{Measurement of the ionization response of amorphous
  selenium with 122 keV \ensuremath{\gamma} rays}},}\ }\href {\doibase
  10.1088/1748-0221/16/06/P06018} {\bibfield  {journal} {\bibinfo  {journal}
  {JINST}\ }\textbf {\bibinfo {volume} {16}},\ \bibinfo {pages} {P06018}
  (\bibinfo {year} {2021})},\ \Eprint {http://arxiv.org/abs/2012.04079}
  {arXiv:2012.04079 [physics.ins-det]} \BibitemShut {NoStop}%
\bibitem [{\citenamefont {Chavarria}\ \emph {et~al.}(2017)\citenamefont
  {Chavarria}, \citenamefont {Galbiati}, \citenamefont {Li},\ and\
  \citenamefont {Rowlands}}]{Chavarria:2016hxk}%
  \BibitemOpen
  \bibfield  {author} {\bibinfo {author} {\bibfnamefont {A.~E.}\ \bibnamefont
  {Chavarria}}, \bibinfo {author} {\bibfnamefont {C.}~\bibnamefont {Galbiati}},
  \bibinfo {author} {\bibfnamefont {X.}~\bibnamefont {Li}}, \ and\ \bibinfo
  {author} {\bibfnamefont {J.~A.}\ \bibnamefont {Rowlands}},\ }\bibfield
  {title} {\enquote {\bibinfo {title} {{A high-resolution CMOS imaging detector
  for the search of neutrinoless double $\beta$ decay in $^{82}$Se}},}\ }\href
  {\doibase 10.1088/1748-0221/12/03/P03022} {\bibfield  {journal} {\bibinfo
  {journal} {JINST}\ }\textbf {\bibinfo {volume} {12}},\ \bibinfo {pages}
  {P03022} (\bibinfo {year} {2017})},\ \Eprint
  {http://arxiv.org/abs/1609.03887} {arXiv:1609.03887 [physics.ins-det]}
  \BibitemShut {NoStop}%
\bibitem [{\citenamefont {Frekers}\ \emph {et~al.}(2016)\citenamefont {Frekers}
  \emph {et~al.}}]{Frekers:2016haa}%
  \BibitemOpen
  \bibfield  {author} {\bibinfo {author} {\bibfnamefont {D.}~\bibnamefont
  {Frekers}} \emph {et~al.},\ }\bibfield  {title} {\enquote {\bibinfo {title}
  {{High energy-resolution measurement of the $^{82}$Se$(^3$He,t$)^{82}$Br
  reaction for double-\ensuremath{\beta} decay and for solar neutrinos}},}\
  }\href {\doibase 10.1103/PhysRevC.94.014614} {\bibfield  {journal} {\bibinfo
  {journal} {Phys. Rev. C}\ }\textbf {\bibinfo {volume} {94}},\ \bibinfo
  {pages} {014614} (\bibinfo {year} {2016})}\BibitemShut {NoStop}%
\end{thebibliography}%

\end{document}